# A brief review of high-entropy oxides in solid oxide fuel cell applications

Yueyuan Gu[a,*], Juan Shi[a], Dilshod Nematov[b], Aoqi Liu[a], Yanru Yin[a], Hailu Dai[c], Lei Bi[a,*]

[a] School of Resource Environment and Safety Engineering, University of South China, Hengyang, 421001, China
[b] S.U. Umarov Physical-Technical Institute of NAST, Dushanbe 734063, Tajikistan
[c] School of Materials Science and Engineering, Yancheng Institute of Technology, Yancheng 224051, China

**Abstract.** Solid oxide fuel cells are efficient energy conversion devices essential to clean energy development, yet their broad application is limited by material challenges, including sluggish oxygen reduction kinetics at intermediate temperatures, electrode instability and vulnerability to contaminants. High-entropy oxides, a novel class of materials characterized by multiple principal elements and high configurational entropy, present a promising approach to overcome these issues via their distinctive "four core effects". This review begins with the fundamentals of high-entropy materials, covering their definition, phase stabilization mechanisms, and relevant descriptors, then systematically reviews their progress as SOFC cathodes, electrolytes, and anodes. Key advances are summarized, and current challenges are analyzed, offering guidance for the design of high-performance and stable high-entropy oxides for solid oxide fuel cells.

# 1. Introduction

Energy and environment are fundamental to sustaining high-quality development in modern society. Solid oxide fuel cells (SOFCs), as a highly efficient and clean energy conversion device, offer robust support for the low-carbon energy transition. Their all-solid-state structure prevents electrolyte loss and corrosion, coupled with advantages such as no need for precious metal catalysts and high fuel flexibility, making SOFCs a prominent research focus [1]. Based on the charge carrier, SOFCs are broadly categorized into oxygen-ion conducting (O-SOFC) and proton-conducting (H-SOFC) solid oxide fuel cells [2]. Structurally, a typical SOFC is composed of a dense electrolyte layer to prevent fuel crossover and short-circuiting sandwiched between two porous electrodes, an anode for fuel oxidation and a cathode for oxygen reduction reaction (ORR) [3]. Lowering the operation temperature of SOFC to an intermediate range (e.g., 500-700 °C) is a prevailing trend, owing to advantages including reduced interconnect and sealant cost, suppressed component degradation, and accelerated start-up and shutdown processes [4-8]. However, it presents a core challenge for the cathode as ORR is thermally activated, and its kinetics slow considerably at lower temperatures, leading to substantial polarization losses [9]. Traditional Co-based cathodes offer high activity but face issues such as cobalt evaporation, significant thermal expansions, and relatively high cost, while Co-free materials still require further performance improvement [10, 11]. Apart from this intrinsic activity issue, conventional cathode materials, which are predominantly Sr/Ba-containing perovskites of the $ABO_3$ or $AA'BB'O_{5+\delta}$ type, suffer from several critical problems. These include surface segregation of Ba/Sr species and susceptibility to poisoning by gaseous impurities (e.g., $CO_2$, $H_2O$), Cr from interconnects, and Si from sealing materials [5, 12-16]. Furthermore, as one of the core components of SOFCs, electrolyte must exhibit high ionic conductivity, negligible electronic conduction, gas tightness, chemical stability across oxidizing and reducing atmospheres, and compatibility with electrodes [17]. Anode materials also face significant challenges, although conventional nickel-based ceramic anodes offer high catalytic activity and low cost, they are susceptible to deactivation through carbon deposition and sulfur poisoning [18].

In recent years, high-entropy materials (HEMs) have attracted significant attention as a novel strategy to address these issues. High-entropy materials (HEMs) are characterized by the formation of a homogeneous solid solution of five or more elements into a single-phase system [19]. This unique atomic-level disorder leads to a high configurational entropy, which stabilizes single-phase structures and gives the four core effects different from traditional materials, high-entropy effect, slow diffusion effect, severe lattice distortion and cocktail effect [20]. The high-entropy effect plays a primary role in stabilizing solid solutions by significantly expanding the solubility limit [21]. Driven by atomic radii and electronegativity differences from multiple elements, high-entropy engineering reinforces severe lattice distortion [22]. The heavily distorted lattice in high-entropy materials significantly raises the energy barriers for atomic migration, leading to the characteristic sluggish diffusion and enhanced phase stability [23]. The cocktail effect is the synergistic interplay of multiple elements across different structural scales, resulting in superior integrated properties.

Herein, this review focuses on the application of high-entropy oxides (HEOs) in SOFCs. While a number of previous reviews have summarized the synthesis methods of high-entropy materials [19, 24-29], the present work systematically surveys recent advances in the SOFC applications. Following an introduction to the fundamental principles and relevant parameters of high-entropy oxides, the review covers progress in cathode, electrolyte, and anode development, with particular attention given to cathode-related studies due to their current research predominance. Current challenges and future research directions are discussed, offering perspectives to advance this rapidly evolving field.

**2. Fundamentals of high-entropy oxides**

The total configuration entropy ($\Delta S_{config}$, also referred to as the mixing entropy $\Delta S_{mix}$) of a random solid solution can be defined by equation 1 [30]. More specifically, for a ABO$_3$ perovskite oxide, the value corresponds to the sum of the configurational entropies of the A-site, B-site, and O-site, as expressed in equation 2 [31]. Similarly, the configurational entropy of spinel and R-P oxides can also be calculated accordingly, and the specific formulas are not given here. It should be noted that there remains ongoing discussion regarding the definition and calculation of configurational entropy in complex oxides. For instance, some studies adopt the configurational entropy metric proposed by Dippo [32, 33], while others do not consider the sum of configurational entropies from both the A- and B-sites. In this paper, for consistency, the total configurational entropy of different sites is used as the criterion for entropy classification.

$$\Delta S_{config} = -R \sum_{i=1}^{n} x_i ln x_i \qquad (1)$$

$$\Delta S_{config} = -R \left[ \left( \sum_{a=1}^{n} x_a ln x_a \right)_{A-site} + \left( \sum_{b=1}^{n} x_b ln x_b \right)_{B-site} + 3 \left( \sum_{c=1}^{n} x_c ln x_c \right)_{O-site} \right] \qquad (2)$$

In the above equation, R is the ideal gas constant, n denotes the number of elements occupying the same ion position. $x_a$, $x_b$ and $x_c$ are the mole fraction of the ions at A, B, and O sites, respectively. It is important to note that this calculation assumes a stoichiometric oxygen content, and any oxygen deficiency is not accounted for in this standard formulation of configurational entropy. It is widely accepted that $\Delta S \geq 1.5R$ means high-entropy materials, while $1.0R \leq \Delta S < 1.5R$ and $\Delta S < 1.0R$ denotes medium-entropy and low-entropy materials, respectively. For a fixed elements number (n) at a given lattice site, the maximum configurational entropy is achieved when all constituent ions are in equal molar fractions. This maximum value depends solely on n, irrespective of the specific elemental species involved [25, 34]. Under the condition of equimolar composition for n elements at a single site, where $x_i = \frac{1}{n}$, Equation 1 can be simplified to Equation 3.

$$\Delta S_{config} = -R \left[ \left( \frac{1}{n} \right) ln \frac{1}{n} \times n \right] = -R ln \frac{1}{n} = R ln n \qquad (3)$$

$$\Delta G = \Delta H - T \Delta S \qquad (4)$$

Based on the Gibbs-Helmholtz formula (see equation 4), a negative change in the Gibbs-free energy change $\Delta G$ is required to stabilize a single-phase solid solution [19]. Where $\Delta H$ and $\Delta S$ are the enthalpy and entropy changes, respectively, and $T$ is the temperature. A higher configurational entropy, according to Equation 4, lowers the

Gibbs free energy change, thereby facilitating the formation of materials at high temperatures [26]. Maria *et al.* [35] found that the transition from a multiphase mixture to a single-phase solid solution in an equimolar (MgO, CoO, NiO, CuO, ZnO) system is endothermic, driven by the entropy increase from cation randomization. Thermodynamic calculations at 875 °C confirmed the configurational entropy contribution (~15 kJ·mol$^{-1}$) exceeded the enthalpy cost (~10 kJ·mol$^{-1}$), establishing entropy as the core driving force. This entropy-driven stabilization is maximized at the equimolar composition, as evidenced by an increase in the transition temperature when deviating from this ratio. In other words, the high configurational entropy, maximized at equimolar compositions, stabilizes single-phase solid solutions by lowering the Gibbs free energy at high temperatures, thereby overcoming a positive enthalpy and enabling synthesis at the minimal feasible temperature. And when the enthalpy change ($\Delta H$) is already negative, the entropy stabilization effect is generally weak. Based on this, high-entropy oxides can be classified into entropy-stabilized and non-entropy-stabilized types [36].

While configurational entropy facilitates single-phase high-entropy materials formation, it is not a deterministic predictor. For perovskite oxides, the Goldschmidt tolerance factor (t, equation 5) offers a more reliable criterion for assessing single-phase formability [37]. For perovskite oxide ABO$_3$, $r_A$, $r_B$ and $r_O$ are the radii of the cations on the A, B and O sites, respectively. When multiple ions occupy the A or B site, the relevant ionic radius for subsequent calculations is the mole-fraction-weighted average. The tolerance factor provides an initial screening for the potential formation of single-phase perovskite high-entropy oxides. The ideal cubic perovskite requires the Goldschmidt tolerance factor (t) equals to 1. A stable perovskite structure forms when 0.78 < t < 1.05 [38]. And a cubic-phase forms at $0.9 \leq t \leq 1.0$, with orthorhombic/rhombohedral phases for t < 0.9, or hexagonal/tetragonal phases for t > 1.0, respectively [39]. According to Tang *et al.*'s [40] report, the formation of a single-phase cubic structure in high-entropy perovskites demands a much narrower tolerance factor range of 0.97 to 1.03, and a second phase will appear when t>1.03. Note that the t value close to the ideal 1.00 is probably necessary but not sufficient for the formation of a single cubic high-entropy perovskite phase [39]. And as shown in Fig. 1a, high-entropy perovskites may exhibit higher crystallographic symmetry with greater cation size differences and smaller Goldschmidt tolerance factors [31]. In addition, the modified tolerance factor ($\tau$), calculated using ionic radii and oxidation states (see equation 6, $n_A$ is the oxidation state of ion A in ABX$_3$ structures), serves as a key criterion for predicting the formation of a pure perovskite phase. To ensure solid solubility and structural stability, $\tau$ should generally remain below 4.18, with lower values favoring the formation of a single phase [41]. For high-entropy perovskites, $\tau$ < 3 serves as a more reliable criterion for screening single-phase materials [42]. Fig. 1b-c illustrate a screening strategy for optimal high-entropy perovskites using the tolerance factor and mixing enthalpy as dual criteria.

$$t = \frac{r_A + r_O}{\sqrt{2}(r_B + r_O)} \quad (5)$$

$$\tau = \frac{r_X}{r_B} - n_A \left[ n_A - \frac{r_A/r_B}{\ln\frac{r_A}{r_B}} \right] \quad (6)$$

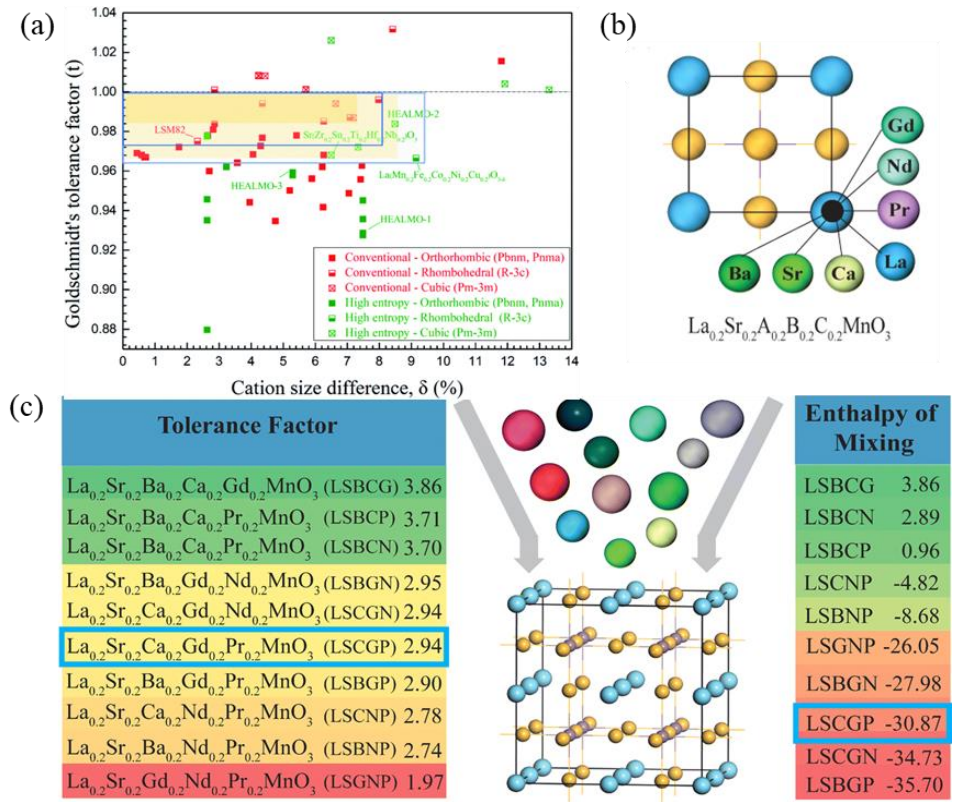

*Figure 1. (a) Correlation of the crystal structure with the Goldschmidt tolerance factor and cation size difference. For t < 1, the yellow shaded areas demonstrate the boundaries of the cubic structure for conventional and high entropy materials and the green boxes for the rhombohedral, reproduced with permission [31]. (b) A-site dopants (Pr, Gd, Nd, Ba, and Ca) considered in ref 42 to form high entropy perovskites of $La_{1-x}Sr_xMnO_{3-\delta}$ (LSM) and (c) Screening of possible high entropy configurations based on tolerance factor and enthalpy of mixing (kJ mol$^{-1}$), where the high entropy configuration in the blue box represents the chosen material LSCGP, reproduced with permission [42].*

In the machine learning analysis by Ma *et al.* [43], the configurational entropy (ΔS) was not identified as a primary predictor for single-phase formation in high-entropy spinel oxides. Instead, the volume deviation of the precursor oxides was revealed as the most critical feature, followed by the deviations in electronegativity and density. A smaller volume deviation of the precursor oxides reflecting closer atomic sizes and favoring uniform structures formation. The ion size difference factor ($\delta$) is derived from the Hume-Rothery solid-solution rule using equation 7 [21, 31]. The A-site cation-size difference ($\delta_{r_A}$) is given here, while the B-site one is omitted as they share a similar calculation formula. It has been reported that for high-entropy alloys, single solid phases tend to form when $\delta <\sim 6.5\%$. For single high-entropy perovskite phases formation, the threshold appears to be much higher, as evidenced by the formation of single-phase high-entropy perovskites when $\delta_{r_B} =13.3\%$ [39]. This means that high-entropy perovskites can accommodate a notably larger difference in ionic sizes. However, as evidenced in LaMnO$_3$ based high entropy perovskite oxides (HEPOs), neither the Goldschmidt tolerance factor (t) nor the cation size difference ($\delta$) can be simply used to predict the phase stability of HEPOs [31]. Beyond size and geometric factors, cation valence also plays a critical role, a higher cation valence difference promotes deviation from a random distribution, driving the system toward a partially ordered structure [44], which could be detrimental to the formation of a homogeneous high-entropy solid solution.

$$\delta_{r_A} = \sqrt{\sum_{i=1}^{n} x_i \left(1 - \frac{r_{A_i}}{\left(\sum_{i=1}^{n} x_i r_{A_i}\right)}\right)^2} \qquad (7)$$

The random occupation of cations with varied ionic radii at lattice sites induces severe lattice distortion in high-entropy materials [45, 46]. This lattice distortion can be quantified by the distortion parameter ($\varepsilon$), which assesses the deviation from ideal cubic symmetry in pseudo-cubic systems (see equation 8) [47].

$$\varepsilon = \left(\frac{1}{3}\right)\left[\left\{\frac{a_{norm}-a_{ps.cubic}}{a_{ps.cubic}}\right\}^2 + \left\{\frac{b_{norm}-a_{ps.cubic}}{a_{ps.cubic}}\right\}^2 + \left\{\frac{c_{norm}-a_{ps.cubic}}{a_{ps.cubic}}\right\}^2\right]^{0.5} \quad (8)$$

$$a_{ps.cubic} = \sqrt[3]{V} \quad (9)$$

$$a_{norm} = \frac{a}{\sqrt{2}\sqrt[3]{V}}, b_{norm} = \frac{b}{\sqrt{2}\sqrt[3]{V}}, c_{norm} = \frac{c}{2\sqrt[3]{V}} \quad (10)$$

$$a_{norm} = \frac{a}{\sqrt{3}\sqrt[3]{V}}, b_{norm} = \frac{b}{\sqrt{3}\sqrt[3]{V}}, c_{norm} = \frac{c}{2\sqrt[3]{V}} \quad (11)$$

Where $a_{ps.cubic}$ indicates the pseudo-cubic lattice parameter, which can be derived use equation 9. The parameters of $a_{norm}$, $b_{norm}$ and $c_{norm}$ are the normalized pseudo-cubic lattice parameters of a symmetry cell. For orthorhombic or cubic unit cells, their volumes are four times larger than those of the pseudo-cubic one, thus the $a_{norm}$, $b_{norm}$ and $c_{norm}$ can be calculated use equation 10. And the volume of rhombohedral unit cell is six times larger than the pseudo-cubic one, so the corresponding parameters can be derived from equation 11 [31]. The parameters of $V$, $a$, $b$ and $c$ are the cell volume and perovskite lattice parameters obtained from XRD refinement. A higher $\varepsilon$ value indicates lower structural symmetry and a more distorted lattice [38]. This lattice distortion induces a disordered stress field that kinetically suppresses cation segregation (e.g., Sr) by hindering its transport and migration [47].

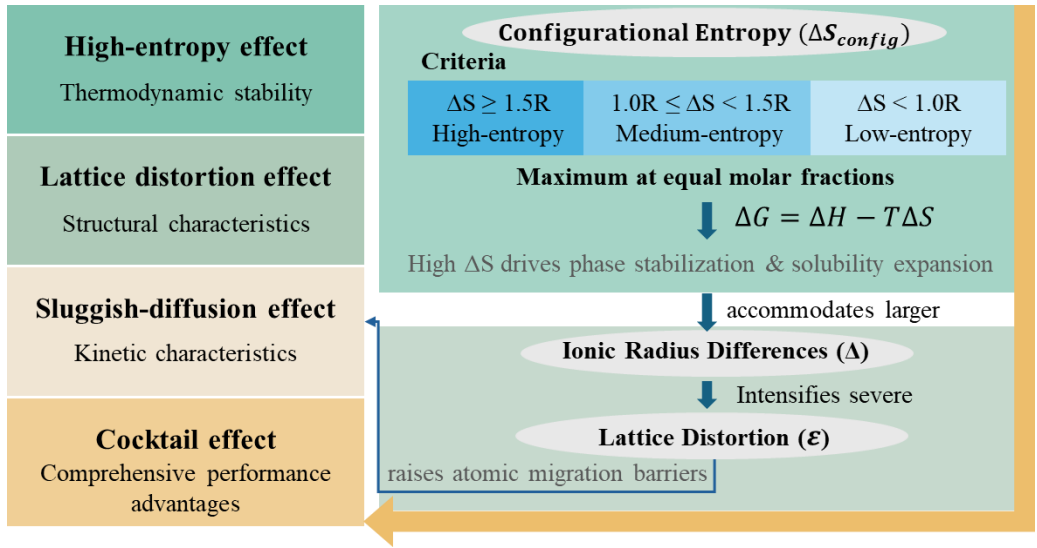

*Figure 2. The four core effects and their interrelationships in high-entropy materials.*

In summary, as illustrated in Fig. 2, the four core effects of high-entropy materials are interrelated and can be systematically understood through key parameters. Essentially, high configurational entropy serves as the key thermodynamic driver for the formation of single-phase solid solutions. The elevated entropy stabilizes the single-phase structure and expands the solubility limit, which enables the accommodation of larger ionic radius differences and consequently leads to severe lattice distortion. This distortion raises the energy barriers for atomic migration, leading to the sluggish diffusion effect. And these effects work synergistically, yielding superior integrated properties known as the cocktail effect.

## 3. Application for SOFC cathodes

Based on the fundamental principles of high-entropy materials discussed above, their application in SOFC cathodes has emerged as a promising strategy to overcome the limitations of conventional cathode materials. Particularly the severe lattice distortion and sluggish diffusion, which offers high-entropy oxide cathodes with exceptional structure stability and suppressed cation segregation, while the cocktail effect offers the potential for comprehensive performance enhancement. The applications of high-entropy oxide cathodes in O-SOFCs and H-SOFCs are reviewed in the following part, respectively.

**3.1 HEOs for O-SOFC cathodes**

Ma *et al.* [43, 48-51] investigated a series of equimolar high-entropy spinel cathode materials, including the A/B-site mixed high entropy $(Fe_{0.2}Mn_{0.2}Co_{0.2}Ni_{0.2}Zn_{0.2})_3O_4$ and $Mn_{0.6}Fe_{0.6}Co_{0.6}Cu_{0.6}Mg_{0.6}O_4$, the A-site high-entropy $(Mg_{0.2}Fe_{0.2}Co_{0.2}Ni_{0.2}Cu_{0.2})Fe_2O_4$, and the B-site high-entropy $Ni(Fe_{0.2}Mn_{0.2}Co_{0.2}Cr_{0.2}Ni_{0.2})_2O_4$. All materials exhibited abundant oxygen vacancies, which is beneficial for enhancing ORR activity. And the high configurational entropy suppressed secondary phases formation, lattice distortion and disordered stress fields slowed atomic diffusion, ensuring excellent stability. The electrochemical performance of these high-entropy spinel cathodes in O-SOFCs is compared in Fig. 3. Among them, the $(Mg_{0.2}Fe_{0.2}Co_{0.2}Ni_{0.2}Cu_{0.2})Fe_2O_4$-GDC composite cathode achieved a peak power density of 1063.94 mW/cm$^2$ at 800 °C, while maintaining its microstructure throughout 240 h of operation at 750 °C and 0.5 V with negligible performance degradation [51]. The cobalt-free $(NiFeCrZnMg)_3O_4$ [52] and non-equimolar $(Mn_{3/11}Fe_{3/11}Co_{3/11}Ni_{1/11}Cu_{1/11})_3O_4$[53] high-entropy cathodes also exhibited increased oxygen vacancy concentration and enhanced oxygen kinetics, demonstrating the potential of entropy-driven cationic disorder in SOFC cathodes design.

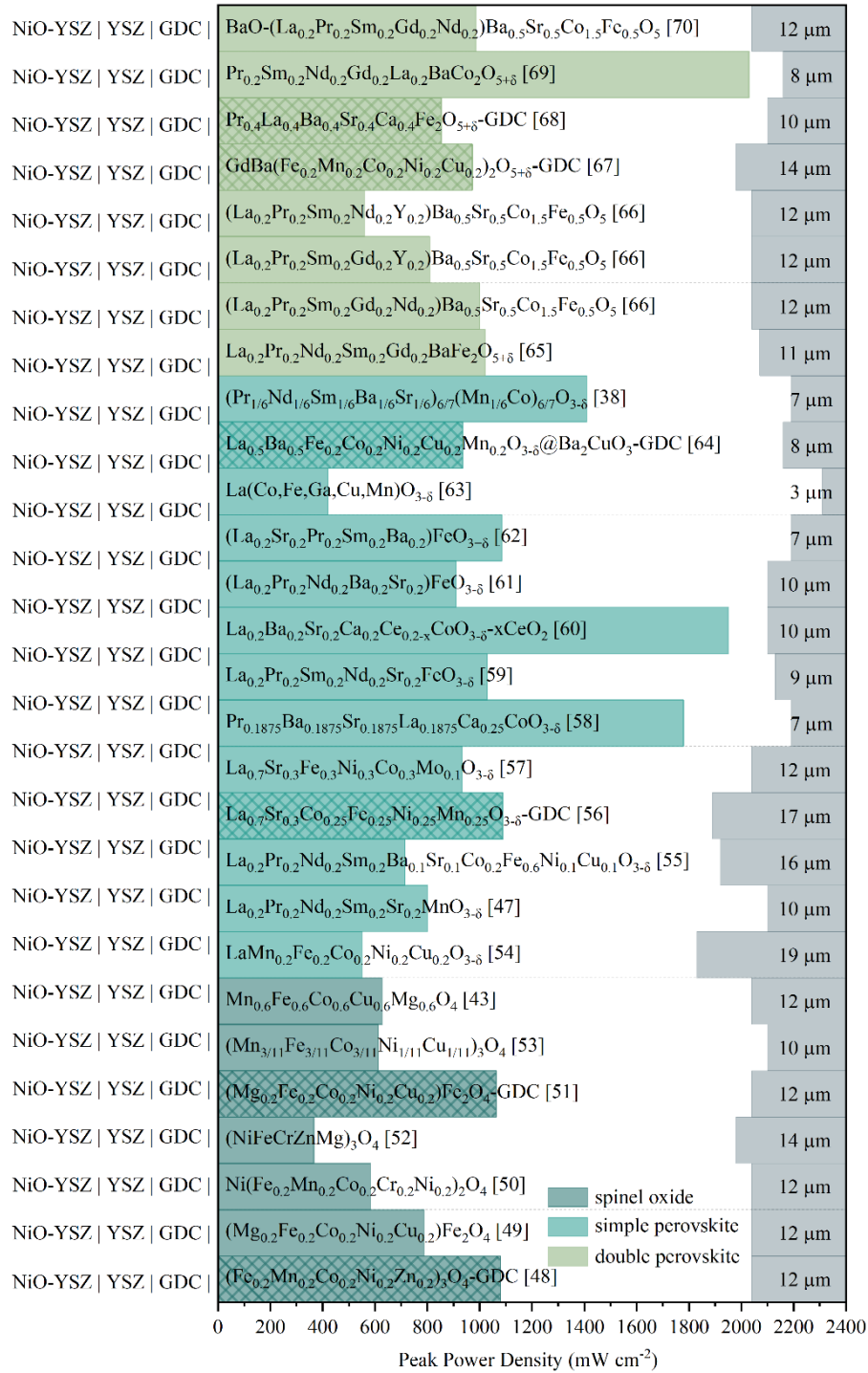

*Figure 3. Horizontal bar comparison of peak power density at 800 °C and corresponding electrolyte thickness for O-SOFCs with different high-entropy cathodes under the same anode supported configuration. Hatched bars denote cells using composite cathodes with gadolinium-doped ceria (GDC). Colored blocks correspond to spinel oxide, simple perovskite, and double perovskite cathodes (see legend). Data*

Sr segregation on perovskite cathodes severely degrades the ORR activity and durability of SOFC. Herein, Ling *et al.* designed a Sr-free B-site high-entropy cathode, LaMn$_{0.2}$Fe$_{0.2}$Co$_{0.2}$Ni$_{0.2}$Cu$_{0.2}$O$_{3-\delta}$ [54], and an A-site high-entropy cathode, La$_{0.2}$Pr$_{0.2}$Nd$_{0.2}$Sm$_{0.2}$Sr$_{0.2}$MnO$_{3-\delta}$ [47]. In the A-site high-entropy system, Sr segregation is effectively suppressed as the introduced lattice distortion generates a disordered stress field around Sr that limits Sr migration. This suppressed Sr segregation by high-entropy strategy is further validated in Fe-based high-entropy perovskites, such as La$_{0.2}$Pr$_{0.2}$Sm$_{0.2}$Nd$_{0.2}$Sr$_{0.2}$FeO$_{3-\delta}$ [59] and (La$_{0.2}$Nd$_{0.2}$Sm$_{0.2}$Sr$_{0.2}$Ba$_{0.2}$)Co$_{0.2}$Fe$_{0.8}$O$_3$ [71], both featuring an equimolar five-component in A-site. Moreover, this suppression of Sr segregation also substantially enhances resistance to $CO_2$, as evidenced in (La$_{0.2}$Sr$_{0.2}$Pr$_{0.2}$Sm$_{0.2}$Ba$_{0.2}$)FeO$_{3-\delta}$ [62]. While these A-site high-entropy designs effectively mitigate Sr segregation, a more direct strategy is to completely eliminate alkaline earth metals (e.g., Sr, Ba) from the cathode. By removing these primary reactive sites, such alkali-free perovskites are theoretically expected to exhibit superior tolerance against both Cr-poisoning and $CO_2$-induced deposition. Klaudia *et al.* [63] designed a series of alkali-free B-site high-entropy La(Co, Fe, Ga, X$_1$, X$_2$)O$_3$ (X$_1$, X$_2$ = Cu, Mg, Mn, Ni) perovskite cathodes, only four of the six obtained single-phase compositions, revealing a need for better understanding of phase formation in high-entropy systems. Nevertheless, the compositional flexibility of high-entropy perovskites allows for the rational selection of elements to balance key properties. Recent research highlights the critical role of A-site composition in optimizing both the

ORR kinetics and the thermal expansion behavior [72]. Li *et al.* [73] demonstrated the rational optimization of A-site high-entropy perovskites by retaining conductive alkaline-earth elements Sr/Ba while screening for highly active elements. Their univariate analysis established the following activity trend for A-site cations: Nd < Sm < Ba < Y. Consequently, the optimal composition $(La_{0.2}Sr_{0.2}Pr_{0.2}Y_{0.2}Ba_{0.2})Co_{0.2}Fe_{0.8}O_{3-\delta}$, which incorporates the high-activity Y and Ba showed the best performance, validating the proposed design strategy.

In some cases, the atomic size effect can induce an in-situ self-construction of heterostructures, which potentially extends the triple-phase boundaries and increases the number of active sites. Coupled with the high-entropy effect that suppresses elemental segregation, such a microstructure enables the cathode material to achieve a synergistic combination of high activity and long-term stability [64]. A representative example is the in-situ construction of an A-site high-entropy perovskite composite cathode, $La_{0.2}Ba_{0.2}Sr_{0.2}Ca_{0.2}Ce_{0.2-x}CoO_{3-\delta}-xCeO_2$ (LBSCCC–CeO$_2$) [60]. This material achieved a remarkable peak power density of 1.95 W cm$^{-2}$ at 800 °C. Simultaneously, it demonstrated exceptional operational stability, with an area-specific resistance (ASR) degradation rate as low as 0.00021 $\Omega\cdot cm^2\cdot h^{-1}$ during 100 h test at 750 °C in air. This performance validates the potential of combining entropy engineering with interface design.

Beyond equimolar designs, non-equimolar high-entropy perovskites offer a more flexible and tunable compositional space. For instance, conventional cathodes can be transformed by multi-element doping, such as high-entropy

$La_{0.7}Sr_{0.3}Fe_{0.3}Ni_{0.3}Co_{0.3}Mo_{0.1}O_{3-\delta}$ derived from conventional $La_{0.7}Sr_{0.3}FeO_{3-\delta}$ (LSF) [57], $La_{0.7}Sr_{0.3}Co_{0.25}Fe_{0.25}Ni_{0.25}Mn_{0.25}O_{3-\delta}$ [56]. An even more complex strategy involves the simultaneous incorporation of five distinct elements on both A- and B-sites, as exemplified by $La_{0.2}Pr_{0.2}Nd_{0.2}Sm_{0.2}Ba_{0.1}Sr_{0.1}Co_{0.2}Fe_{0.6}Ni_{0.1}Cu_{0.1}O_{3-\delta}$ [55]. This non-equimolar high-entropy design allows for selective adjustment of compositions to optimize specific properties while retaining the fundamental benefits of high configurational entropy. For example, the Ca-rich $Pr_{0.1875}Ba_{0.1875}Sr_{0.1875}La_{0.1875}Ca_{0.25}CoO_{3-\delta}$ exhibits an O $p$-band center at -3.26 eV, which is closer to the Fermi level compared to -3.29 eV in equimolar $Pr_{0.2}Ba_{0.2}Sr_{0.2}La_{0.2}Ca_{0.2}CoO_{3-\delta}$ [58, 74]. Consequently, the ORR kinetics are effectively enhanced, the peak power density at 750 °C increased from 0.78 W cm$^{-2}$ to 1.14 W cm$^{-2}$.

$(La_{0.2}Pr_{0.2}Nd_{0.2}Sm_{0.2}Gd_{0.2})_2CuO_4$ was the first reported high-entropy $A_2BO_4$-type cathode for SOFCs, it is reported that entropy stabilization suppresses element segregation and enhances ORR activity, achieving an area-specific resistance of 0.52 Ω cm$^2$ and a peak power density of 528 mW cm$^{-2}$ at 700 °C [75]. Wei *et al.* [76, 77] synthesized a B-site medium entropy $GdBa(Mn_{0.2}Fe_{0.2}Co_{1.2}Ni_{0.2}Cu_{0.2})O_{5+\delta}$ double perovskite cathode, the lattice distortion effectively suppresses $Ba^{2+}$ diffusion while simultaneously inducing the in-situ formation of $BaCoO_{3-\delta}$ nanoparticles, exhibiting excellent electrochemical activity and robust stability against $CO_2$ and Cr poisoning. In contrast, the B-site equimolar high-entropy $GdBa(Fe_{0.2}Mn_{0.2}Co_{0.2}Ni_{0.2}Cu_{0.2})_2O_{5+\delta}$ [67] did not exhibit this in-situ $BaCoO_{3-\delta}$ exsolution phenomenon. This may be attributed to the increased configurational entropy, which promotes a more homogeneous cation distribution of B-site, coupled with the significantly reduced Co content in the equimolar composition. Similarly, in the A-site equimolar high-entropy $(La_{0.2}Pr_{0.2}Nd_{0.2}Sm_{0.2}Gd_{0.2})BaCo_2O_{5+\delta}$ cathode with high Ba and Co content, reduced surface Ba segregation and the formation of active $BaCoO_{3-\delta}$ nanoparticles were observed, which enhanced its stability in a $CO_2$-containing atmosphere and ORR activity [78]. Conventional double perovskites (e.g., $PrBa_{0.8}Ca_{0.2}Co_2O_{5+\delta}$, PBCC) typically suffer from severe $BaO/Co_3O_4$ segregation, which can be efficiently suppressed by the A-site entropy strategy, as validated in medium entropy $PrBa_{0.5}Sr_{0.25}Ca_{0.2}Ce_{0.05}Co_2O_{5+\delta}$ [79] and $Pr_{0.6}La_{0.1}Nd_{0.1}Sm_{0.1}Gd_{0.1}Ba_{0.8}Ca_{0.2}Co_2O_{5+\delta}$ [80] cathodes. Moreover, the conventional PBCC cathode faces challenges from Co spin state transition-induced lattice expansion, and the variable valence state of Co during redox cycling can lead to performance degradation. The B-site high-entropy engineering offers a solution by reducing Co content, improving thermal expansion compatibility with the electrolyte, and introduce synergistic effects among multiple elements. The B-site high-entropy $PrBa_{0.8}Ca_{0.2}Fe_{0.4}Co_{0.4}Ni_{0.4}Cu_{0.4}Zn_{0.4}O_{6-\delta}$ achieved a peak power density of 1.42 W cm$^{-2}$ at 700 °C, underscoring the efficacy of this strategy [81].

Zheng *et al.* [66] synthesized three high-entropy double perovskites by introducing equimolar mixtures of different rare-earth elements (La, Pr, Sm, Gd, Nd, Y) into the A-site of $LaBa_{0.5}Sr_{0.5}Co_{1.5}Fe_{0.5}O_5$ (LBSCF), namely as LPSGNBSCF, LPSGYBSCF, and LPSNYBSCF. The optimal LPSGNBSCF cathode achieved a power density of 1000 mW cm$^{-2}$ at 800 °C. Furthermore, infiltrating a BaO protective layer onto the LPSGNBSCF cathode can effectively suppressed Sr segregation and mitigated Cr deposition, significantly enhancing its resistance to Cr poisoning [70]. Using the same equimolar A-site composition (La, Pr, Nd, Sm, Gd), the high-entropy cathode La$_{0.2}$Pr$_{0.2}$Nd$_{0.2}$Sm$_{0.2}$Gd$_{0.2}$BaFe$_2$O$_{5+\delta}$ (LPNSGBF) achieved a similarly peak power density of 1020.69 mW cm$^{-2}$ at 800 °C, and the entropy effect suppressed Ba segregation, resulting in a low Cr-poisoning degradation rate of 0.17% h$^{-1}$ over 100 h [65]. Furthermore, Applying the same A-site high-entropy composition (La, Pr, Nd, Sm, Gd) to a Co-based system yielded an active and contaminants-tolerant high-entropy cathode, Pr$_{0.2}$Sm$_{0.2}$Nd$_{0.2}$Gd$_{0.2}$La$_{0.2}$BaCo$_2$O$_{5+\delta}$ exhibited an exceptional peak power density of 2.03 W cm$^{-2}$ at 800 °C [69], significantly outperforming its Fe-based materials. These results demonstrate the universal effectiveness of the A-site high-entropy design. The high-entropy strategy stabilizes the crystal lattice to mitigate degradation, while the optimization of B-site elements enhances catalytic activity, thereby achieving a simultaneous improvement in both catalytic performance and stability. The cobalt-free high-entropy double perovskite Pr$_{0.4}$La$_{0.4}$Ba$_{0.4}$Sr$_{0.4}$Ca$_{0.4}$Fe$_2$O$_{5+\delta}$ exemplifies the performance-stability synergy from the high-entropy effect, achieving a comparable peak power density (0.853 W cm$^{-2}$) to the cobalt-based PrBa$_{0.5}$Sr$_{0.5}$Co$_{1.5}$Fe$_{0.5}$O$_5$ (0.92 W cm$^{-2}$) while significantly improving durability, exhibiting a reduced 100 h degradation rate of 12.4%/100 h compared to 6.7%/100 h [68].

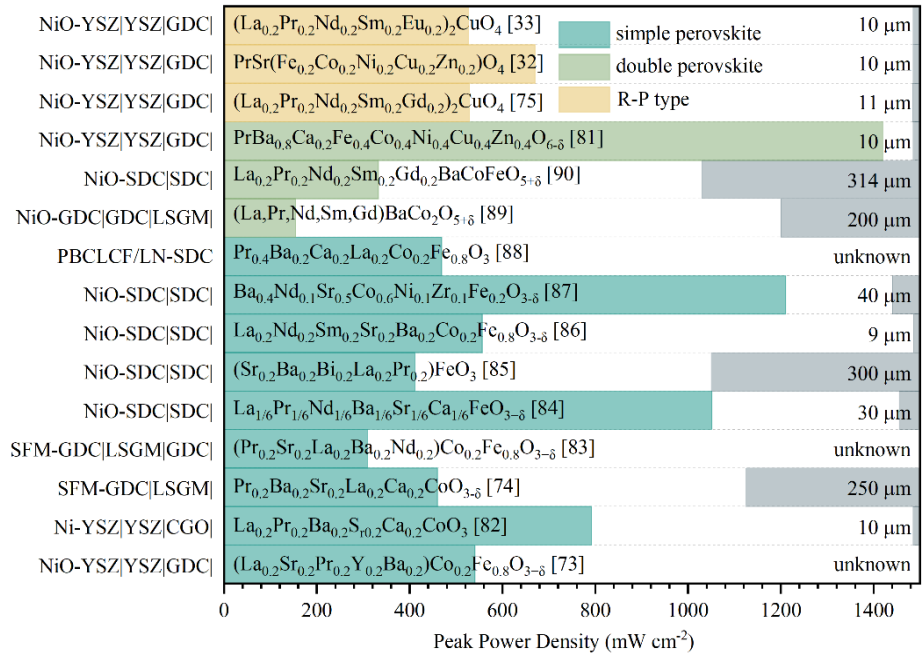

*Figure 4. Horizontal bar comparison of peak power density at 700 °C and corresponding electrolyte thickness for O-SOFCs with different high-entropy cathodes under varied half-cell configuration. Colored blocks correspond to simple perovskite, double perovskite, and R-P type cathodes (see legend). Data from references indicated.*

Juliusz et al. [91] first introduced Cr intentionally into the B-site of a high-entropy cathode, synthesizing $La_{1-x}Sr_x(Co, Cr, Fe, Mn, Ni)O_{3-\delta}$ (x= 0, 0.1, 0.2, 0.3). The results show that the Sr solubility limit in this system is x = 0.3, beyond which Sr and Cr segregation occurs, forming $Sr_3(CrO_4)_2$ impurities. Based on this, they hypothesized that incorporating Cr into the B-site might intrinsically enhance the cathode's resistance to Cr poisoning. However, this hypothesis remains speculative, as subsequent experimental validation has not been reported. Liu *et al.* [82] and He *et al.* [74] independently reported high-entropy cobalt-based perovskite cathodes featuring the same five-component equimolar A-site composition (La, Pr, Ba, Sr, Ca). Both studies observed improved stability, thereby providing mutual validation for the structural stabilization effect of high configurational entropy. As shown in Fig. 4, due to distinct half-cell configurations, particularly electrolyte thicknesses of 10 μm [82] and 250 μm [74], their measured peak power densities differed substantially, reaching 792 mW cm$^{-2}$ and 460 mW cm$^{-2}$ at 700 °C, respectively. Therefore, when evaluating and selecting cathode materials from reported data, factors such as cell design and electrolyte thickness, which are fundamental determinants of the total cell resistance, must be carefully considered alongside the material's intrinsic properties. The relatively low peak power densities observed for the electrolyte-supported cells [85, 89, 90] in Fig. 4 do not imply low ORR activity of these high-entropy cathodes. In fact, given their thick electrolytes, the performance is already significant and can be further enhanced through electrolyte thinning and structure optimization [42].

High-entropy engineering does not always concurrently enhance both activity and stability, a trade-off may exist. For instance, Zheng et al. [86] reported that the A-site high-entropy $La_{0.2}Nd_{0.2}Sm_{0.2}Sr_{0.2}Ba_{0.2}Co_{0.2}Fe_{0.8}O_{3-\delta}$ cathode showed slightly lower initial peak power density than $La_{0.6}Sr_{0.4}Co_{0.2}Fe_{0.8}O_{3-\delta}$ (LSCF), but exhibited much better Cr-poisoning tolerance, with peak power density declining only from 556 to 537 mW·cm$^{-2}$ after 40 h of Cr exposure, compared to a sharp drop from 583 to 390 mW·cm$^{-2}$ for LSCF. The similarly composed high-entropy oxide $(Pr_{0.2}Sr_{0.2}La_{0.2}Ba_{0.2}Nd_{0.2})Co_{0.2}Fe_{0.8}O_{3-\delta}$ [83] also demonstrated limited ORR activity (see Fig. 4). Considering the excellent ORR activity of LSCF and the stabilization effect of high entropy, Han et al. [92] proposed an entropy-assisted surface engineering strategy by impregnating LSCF with La$_{0.2}$Pr$_{0.2}$Nd$_{0.2}$Sm$_{0.2}$Gd$_{0.2}$)$_{0.2}$Ce$_{0.8}$O$_{2-\delta}$. The composite cathode exhibited high ORR activity and strong Cr-poisoning resistance, achieving a peak power density of 1564.85 mW·cm$^{-2}$ at 750 °C and showing only 0.11 %·h$^{-1}$ degradation at 500 mA·cm$^{-2}$ under Cr-containing atmosphere for 100 h. This work demonstrates a promising new approach for designing durable high-performance cathodes.

Fu *et al.* [84] synthesized A-site high-entropy cobalt-free perovskite nanofiber La$_{1/6}$Pr$_{1/6}$Nd$_{1/6}$Ba$_{1/6}$Sr$_{1/6}$Ca$_{1/6}$FeO$_{3-\delta}$ cathodes via electrospinning. This multi-component doping at the A-site induces lattice distortion and enhances oxygen-vacancy concentration, resulting in a low polarization resistance of 0.041 Ω·cm$^2$ at 750 °C and a high peak power density of 1.439 W·cm$^{-2}$. The A-site elements in the aforementioned high-entropy perovskite SOFC cathodes are typically selected from rare-earth and alkaline-earth metals, this trend also applies to double perovskites such as (La,Pr,Nd,Sm,Gd)BaCo$_2$O$_{5+\delta}$ [89], La$_{0.2}$Pr$_{0.2}$Nd$_{0.2}$Sm$_{0.2}$Gd$_{0.2}$BaCoFeO$_{5+\delta}$ [90]. Beyond these, Bi can also occupy the A-site, as exemplified by the high-entropy (Sr$_{0.2}$Ba$_{0.2}$Bi$_{0.2}$La$_{0.2}$Pr$_{0.2}$)FeO$_3$ [85], which demonstrates suppressed Sr segregation, improved chemical compatibility and CO$_2$ tolerance. Based on high-throughput calculation, Li *et al.* [87] screened both the optimal doping elements and ratios from 13 candidate species, selecting Nd, Ni, and Zr as the most favorable dopants for Ba$_{0.5}$Sr$_{0.5}$Co$_{0.8}$Fe$_{0.2}$O$_{3-\delta}$ (BSCF). The high-entropy Ba$_{0.4}$Nd$_{0.1}$Sr$_{0.5}$Co$_{0.6}$Ni$_{0.1}$Zr$_{0.1}$Fe$_{0.2}$O$_{3-\delta}$ shows excellent electronic and ionic conductivity, strong resistance to Cr/CO$_2$ poisoning, and good compatibility with both SDC and

BZCY electrolytes. Single cells with this cathode achieved high power densities of 1.21 W·cm$^{-2}$ (oxide-ion) and 0.63 W·cm$^{-2}$ (proton) at 700 °C, respectively.

**3.2 HEOs for H-SOFC cathodes**

Xu et al. [93] first applied a high-entropy spinel oxide Fe$_{0.6}$Mn$_{0.6}$Co$_{0.6}$Ni$_{0.6}$Cr$_{0.6}$O$_4$ as a cathode for H-SOFCs. This material exhibits a pure spinel structure, homogeneous elemental distribution, and excellent CO$_2$ stability. First-principles calculations indicate that, compared to its single-component counterparts, this material exhibits lower O$_2$ adsorption energy and an O-p band center closer to the Fermi level, achieving a peak power density of 1052 mW·cm$^{-2}$ at 700 °C. Liu *et al.* [94] synthesized an A-site six-component high-entropy perovskite Pr$_{1/6}$La$_{1/6}$Nd$_{1/6}$Ba$_{1/6}$Sr$_{1/6}$Ca$_{1/6}$CoO$_{3-\delta}$ as a bifunctional air electrode for reversible proton-conducting electrochemical cells. This material exhibits triple conductivity (electronic/oxygen-ionic/protonic) and enabled the cell to achieve a peak power density of 1.09 W·cm$^{-2}$ at 600 °C. By replacing the dry-pressed electrolyte with a spin-coated layer, the electrolyte thickness was reduced from 19 μm to ~6.5 μm, further raising the peak power density to 1.21 W·cm$^{-2}$. Similarly, a five-component A-site high-entropy cathode with the same elemental set except for the absence of Nd, Pr$_{0.2}$Ba$_{0.2}$Sr$_{0.2}$La$_{0.2}$Ca$_{0.2}$CoO$_{3-\delta}$ [95] also delivered excellent dual-mode electrochemical performance, achieving a comparable peak power density of 1.16 W·cm$^{-2}$ at 600 °C, further confirming the general applicability of A-site high-entropy design in enhancing both electrode activity and stability. The B-site high-entropy perovskite BaCo$_{0.2}$Fe$_{0.2}$Zr$_{0.2}$Sn$_{0.2}$Pr$_{0.2}$O$_{3-\delta}$ has also been employed as an air electrode in reversible protonic ceramic cells, exhibiting triple conductivity and simultaneously enhancing both electrode catalytic activity and structural stability [96].

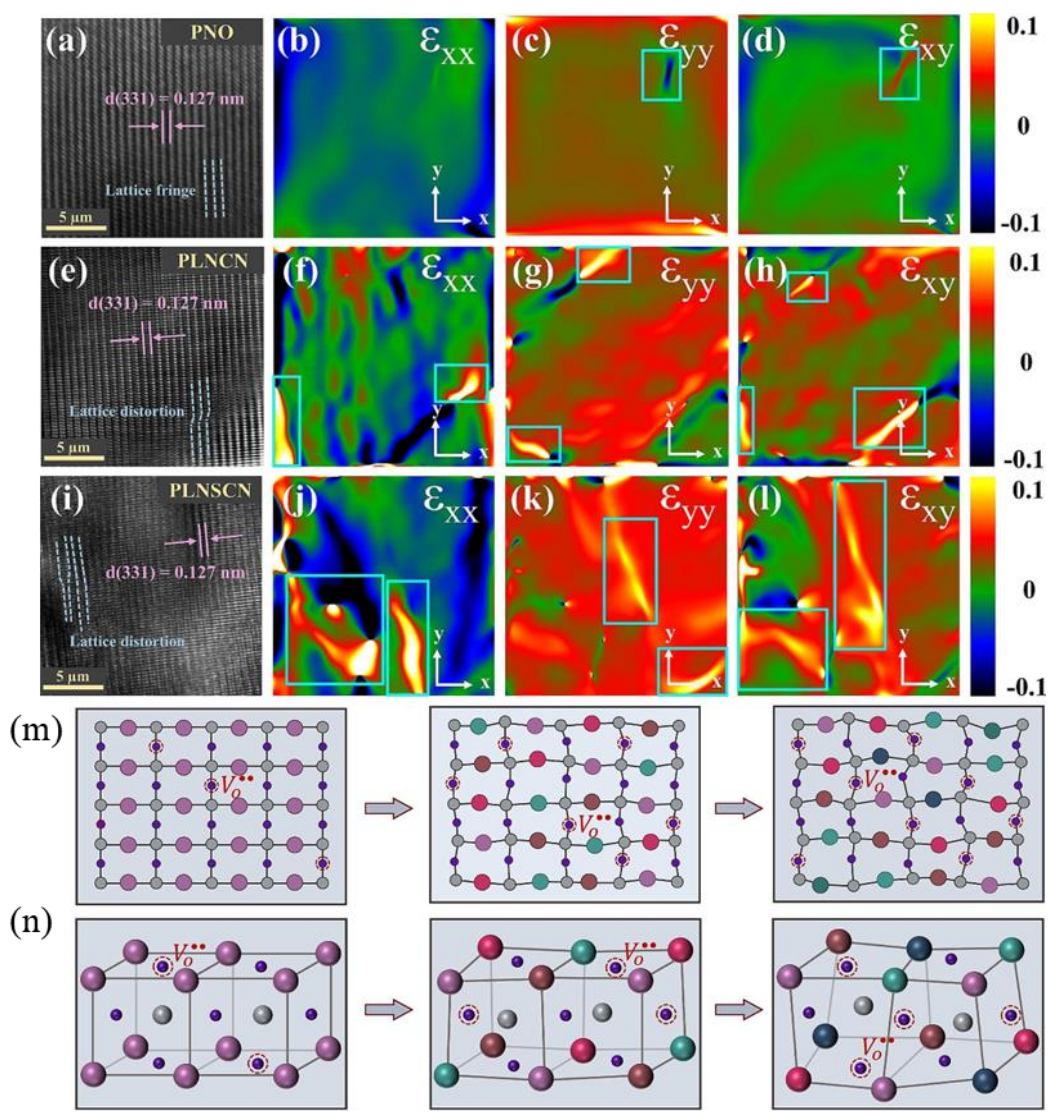

Figure 5. HR-TEM images and corresponding atomic strain distribution of (a)-(d) $Pr_2NiO_{4+\delta}$ (PNO), (e)-(h) $(Pr_{0.25}La_{0.25}Nd_{0.25}Ca_{0.25})_2NiO_{4+\delta}$ (PLNCN), and (i)-(l) $(Pr_{0.2}La_{0.2}Nd_{0.2}Sr_{0.2}Ca_{0.2})_2NiO_{4+\delta}$ (PLNSCN). (m) Planar diagram of the relationship between lattice distortion processes and oxygen vacancies. (n) Three-dimensional schematic diagram of the relationship between lattice distortion processes and oxygen vacancies, reproduced with permission [22].

The Co-free multi-doped RP oxide $Pr_3Ni_{1.5}Cu_{0.3}Nb_{0.05}Ta_{0.05}Zr_{0.05}Y_{0.05}O_{7-\delta}$ exhibits reduced TEC and improved electrolyte compatibility, achieving a peak power density of 1.09 W·cm$^{-2}$ at 600 °C [97]. The B-site equimolar K$_2$NiF$_4$-type cathode La$_{1.2}$Sr$_{0.8}$Mn$_{0.2}$Fe$_{0.2}$Co$_{0.2}$Ni$_{0.2}$Cu$_{0.2}$O$_{4+\delta}$ achieves 1126 mW·cm$^{-2}$ at 600 °C, with its high-entropy design introducing abundant oxygen vacancies, lowering ion-migration barriers, and enhancing stability against CO$_2$ and moisture [98]. Kang et al. [22] systematically tuned the configurational entropy in R-P perovskite cathodes, comparing the low-entropy Pr$_2$NiO$_{4+\delta}$, medium-entropy (Pr$_{0.25}$La$_{0.25}$Nd$_{0.25}$Ca$_{0.25}$)$_2$NiO$_{4+\delta}$, and high-entropy (Pr$_{0.2}$La$_{0.2}$Nd$_{0.2}$Sr$_{0.2}$Ca$_{0.2}$)$_2$NiO$_{4+\delta}$ compositions. With increasing entropy, the crystal structure transforms from orthorhombic to tetragonal, accompanied by enhanced lattice distortion and dislocation (see Fig. 5). The resulting stress-strain fields facilitate the formation of more oxygen vacancies. Consequently, the peak power density increases monotonically with configurational entropy, due to improved oxygen-vacancy concentration, hydration capability, and triple conductivity. The similarly tetragonal-structured La$_{0.4}$Pr$_{0.4}$Nd$_{0.4}$Ba$_{0.4}$Sr$_{0.4}$NiO$_{4+x}$, which replaces Ca with Ba compared to the above composition, exhibits an O-p band center close to the Fermi level at -1.44 eV and delivers an unprecedented peak power density of 2790 mW·cm$^{-2}$ at 700 °C, setting a new performance benchmark for R-P oxide-based cathodes (see Fig. 6) [99]. These results demonstrate that, under similar structural frameworks and configurational entropy levels, the specific choice of constituent elements plays a crucial role in determining the ultimate electrochemical performance of cathode materials.

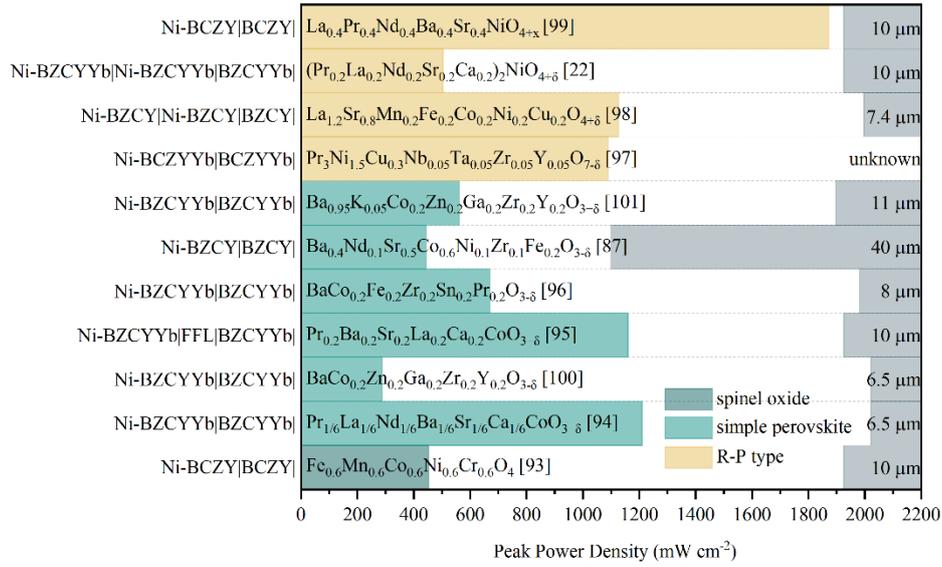

*Figure 6. Horizontal bar comparison of peak power density at 600 °C and corresponding electrolyte thickness for H-SOFCs with different high-entropy cathodes under varied half-cell configurations. Colored blocks correspond to spinel oxide, simple perovskite, and R-P type perovskite cathodes (see legend). Data from references indicated.*

### 3.3 medium entropy oxides for SOFC cathodes

The configurational entropy of $Sr(Fe_\alpha Ti_\beta Co_\gamma Mn_\zeta)O_{3-\delta}$ was tuned via equimolar (SFTCM25) and non-equimolar (SFTCM5221) B-site design. Higher entropy effectively suppresses Sr surface segregation, reduces thermal expansion, enhances oxygen-ion transport, and improves long-term stability. However, these benefits are achieved at the expense of reduced electron conductivity and slightly diminished electrocatalytic activity.[102]. Zhang *et al.* [103] reported a cobalt-free medium-entropy perovskite, $Bi_{0.5}Sr_{0.5}Fe_{0.85}Nb_{0.05}Ta_{0.05}Sb_{0.05}O_{3-\delta}$. The enhanced configurational entropy results in a low polarization resistance of 0.095 Ω cm², a power density of 900 mW cm$^{-2}$ at 700 °C, and excellent $CO_2$ tolerance owing to a more negative average bonding energy. Gao et al. [104, 105, 107] designed a B-site medium-entropy perovskite cathode $Sr_{0.95}Co_{0.5}Fe_{0.2}Ti_{0.1}Ta_{0.1}Nb_{0.1}O_{3-\delta}$ with exceptional $CO_2$ and Cr poisoning tolerance through entropy engineering, and further enhanced its electrochemical performance and Cr resistance via controlled Sr-deficiency optimization.

Recent studies have prompted a critical re-evaluation of the role of entropy in perovskite air electrode design. While medium- and high-entropy configurations undeniably enhance phase stability and mitigate issues like cation segregation, emerging evidence suggests that the intentional selection of specific dopant elements may contribute more decisively to electrocatalytic activity and operational stability than the mere pursuit of high configurational entropy itself [108]. As shown in Fig. 7, the systematic comparisons within a series of A-site engineered perovskites revealed that electrochemical performance does not change monotonically with entropy. Instead, the intrinsic chemical properties of the constituent ions (e.g., $Sr^{2+}$ versus $Cs^+$, see Fig. 5) were found to be the dominant factor governing oxygen reduction kinetics and long-term durability of $Pr_{1/2}Ba_{1/6}Sr_{1/6}Ca_{1/6}CoO_{3-\delta}$ (PBSCC). This indicates that the future of high-performance SOFC cathodes lies in the rational synergy of entropy stabilization and targeted element selection, rather than in the blind pursuit of high configurational entropy. An example is the medium-entropy lithiated oxide cathode $LiCo_{0.25}Fe_{0.25}Mn_{0.25}Ni_{0.25}O_2$. Although its configurational entropy is only moderate, its power output reaches as high as 1803 mW·cm$^{-2}$ at 700 °C, stemming from the synergistic effects of Co, Fe, Mn, and Ni, which together optimize oxygen vacancy formation and the electronic structure [106]. Different dopants offer unique advantages and limitations for cathode performance, so balancing relevant parameters is essential [109]. High-entropy design offers the possibility of coexisting multiple elements, and selecting appropriate elements is crucial for designing high-performance cathodes.

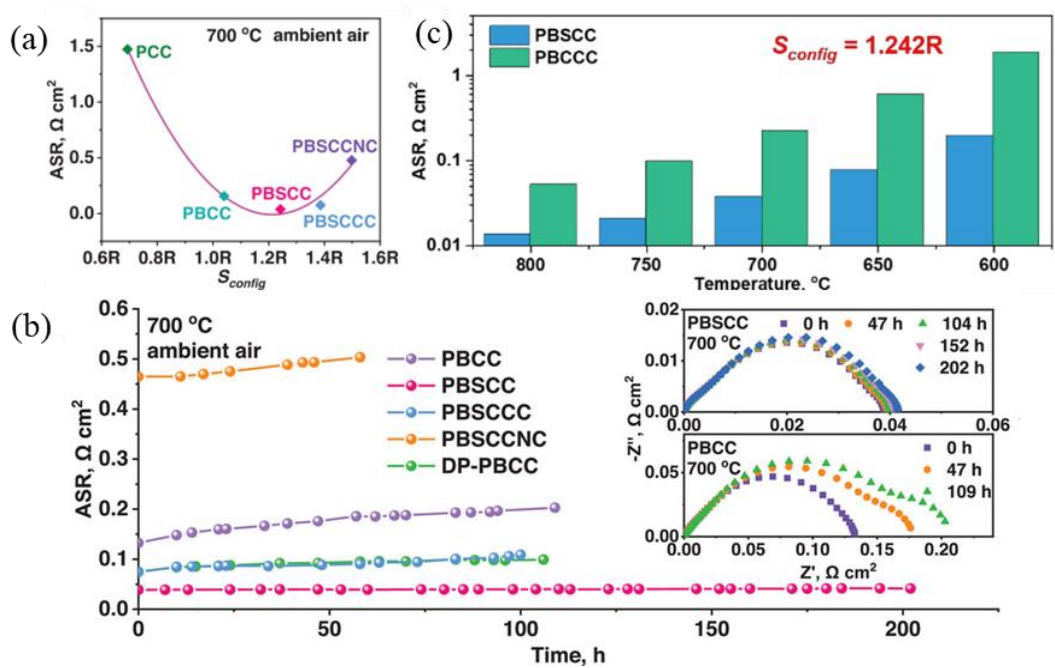

Figure 7. Electrochemical performance of electrode materials with different configurational entropy or doping elements. (a) Relationship between area-specific resistances (ASRs, at 700 °C) and the configurational entropy of developed electrode materials. (b) ASR durability of LSGM-based symmetrical cells with different electrodes in ambient air. The insets are the EIS of PBSCC and PBCC measured under OCV conditions at different times. (c) Comparison of ASRs of these electrodes with the configurational entropy of 1.242 R, reproduced with permission [108].

**4. Application for SOFC electrolytes and anodes**

**4.1 HEOs as SOFC electrolytes**

Conventional oxygen-ion conductors include yttria-stabilized zirconia (YSZ),

gadolinium-doped ceria (GDC), strontium- and magnesium-doped lanthanum gallate (LSGM), and Bi$_2$O$_3$–based electrolytes [110, 111]. Among these materials, 8YSZ is the most mature and widely used, butt its conductivity drops significantly at intermediate-to-low temperatures. GDC exhibits higher activity within this temperature range, but suffers from electronic leakage under low oxygen partial pressures. LSGM has high ionic conductivity, but faces challenges with Ga volatility and reactions with Ni-based anodes [18]. Bi$_2$O$_3$–based electrolytes exhibit relatively high oxygen surface exchange kinetics and O$^{2-}$ conductivity but still face challenges such as poor chemical stability, difficult densification and poor mechanical strength [111]. In recent years, entropy engineering has emerged as a promising strategy for design high-performance electrolytes for SOFCs. Dąbrowa et al. [112] found that Mo-doped ceria-based high-entropy (Ce,Gd,Nd,Sm,Pr)O$_{2-\delta}$ and (Ce,Gd,La,Nd,Pr)O$_{2-\delta}$ can stabilize the disordered fluorite structure and exhibit mixed conduction. Bonnet et al. [113] evaluated the high-entropy fluorite oxides (Hf$_{1/3}$Ce$_{1/3}$Zr$_{1/3}$)$_{1-x}$(Gd$_{1/2}$Y$_{1/2}$)$_x$O$_{1.8}$ and found that its oxygen ion conductivity remains low, only about 4×10$^{-4}$ S·cm$^{-1}$ at 600 °C, significantly lower than the targeted value of ~3×10$^{-2}$ S·cm$^{-1}$ at 600 °C for SOFC application. To overcome this limitation, Li et al. [114] developed a medium-entropy ceria-based electrolyte (Ce$_{0.25}$Sm$_{0.25}$La$_{0.25}$Gd$_{0.25}$)$_2$O$_{3-\delta}$, achieving high ionic conductivity of 0.1534 S·cm$^{-1}$ at 520 °C and demonstrated good proton-oxygen ion mixed conduction, enabling higher fuel cell performance (747 mW cm$^{-2}$ at 520 °C) than that of conventional GDC electrolytes.

Proton-conducting electrolytes are mainly based on BaCeO$_3$ and BaZrO$_3$

perovskites [115]. BaCeO$_3$-based materials exhibit high proton conductivity and good sinterability but poor stability against CO$_2$/H$_2$O, while BaZrO$_3$-based systems exhibit excellent chemical stability and high bulk proton conductivity but suffer from poor sintering and high grain boundary resistance [17, 116]. Gazda *et al.* [117] first examined the proton conduction properties in high-entropy perovskites in 2020, as confirmed by H/D isotope effect, but the conductivity of BaZr$_{0.2}$Sn$_{0.2}$Ti$_{0.2}$Hf$_{0.2}$Y$_{0.2}$O$_{3-\delta}$ is still significantly lower than that of the state-of-art perovskite electrolytes. The triple-doped medium-entropy electrolyte BaCe$_{0.5}$Zr$_{0.2}$Y$_{0.1}$Yb$_{0.1}$Gd$_{0.1}$O$_{3-\delta}$ achieves 10.5 mS·cm$^{-1}$ conductivity of in wet air at 600 °C and shows stable performance in 20% H$_2$O for 200 h [118]. BaSn$_{0.16}$Zr$_{0.24}$Ce$_{0.35}$Y$_{0.1}$Yb$_{0.1}$Dy$_{0.05}$O$_{3-\delta}$ exhibited a conductivity of 8.3 mS·cm$^{-1}$ under 3% H$_2$O humidified air at 600 °C, along with excellent stability in 50% CO$_2$ and boiling water. When integrated into an anode-supported single cell, this material delivered a power density of 318 mW·cm$^{-2}$ at 600 °C [119]. Through systematic screening of the sixth element from Nb, Sn, Gd, and Zn, the equimolar B-site high-entropy BaHf$_{1/6}$Sn$_{1/6}$Zr$_{1/6}$Ce$_{1/6}$Y$_{1/6}$Yb$_{1/6}$O$_{3-\delta}$ achieved an improved total conductivity of 9.2 mS·cm$^{-1}$ at 600 °C. This material also exhibited no significant degradation upon exposure to CO$_2$ and H$_2$O, and delivered an enhanced power density of 0.72 W·cm$^{-2}$ at 600 °C [120]. An innovative approach is to use high-entropy concept to design sintering aids. Using a five-element high-entropy sintering aid (Ni, Cu, Co, Fe, Zn) at 5 mol.% improves the sintering performance of BaCe$_{0.4}$Zr$_{0.4}$Y$_{0.2}$O$_3$ (BCZY) electrolyte. Compared with conventional single-element-doped electrolytes, BaCe$_{0.4}$Zr$_{0.4}$Y$_{0.15}$Ni$_{0.01}$Cu$_{0.01}$Co$_{0.01}$Fe$_{0.01}$Zn$_{0.01}$O$_3$ exhibits enhanced conductivity, indicating that the high-entropy design effectively promotes densification and ionic conduction. Besides, this electrolyte maintains structural stability under $CO_2$ and $H_2O$-containing atmospheres, delivers a power density of 509 mW cm$^{-2}$ at 600 °C, and shows no significant performance degradation during 200 h continuous operation [121]. Yang *et al.* [122, 123] reported entropy-designed B-site multi-component perovskite proton conductors, in which the non-equimolar $BaSn_{0.15}Ce_{0.35}Zr_{0.25}Y_{0.1}Sm_{0.1}Zn_{0.05}O_{3-\delta}$, showed relatively high proton mobility and good stability against $H_2O$ and $CO_2$, but its conductivity remained limited (0.25 mS·cm$^{-1}$ 600 °C). Another composition $BaSn_{0.15}Ce_{0.45}Zr_{0.15}Y_{0.1}Yb_{0.1}Gd_{0.05}O_{3-\delta}$ achieved a higher conductivity of 1.03 mS·cm$^{-1}$ under the same conditions. However, the fuel cell performance of both materials still needs to be verified.

The design of high-entropy electrolytes for proton-conducting SOFCs is primarily focused on $BaCeO_3$–$BaZrO_3$-based solid solutions, where multi-component doping at the B-site has been widely adopted to improve chemical stability under $H_2O/CO_2$-containing atmospheres. It is also noteworthy that while equimolar compositions maximize configurational entropy and benefit sintering densification, non-equimolar designs allow for the adjustment of dominant element content, which may enable a more balanced optimization between proton conductivity and overall performance while maintaining a relatively high entropy configuration.

**4.2 HEOs as SOFC anodes**

The application of high-entropy materials in the anodes of SOFC is an emerging field with limited related reports. When used as SOFC anodes, high-entropy alloys

(HEAs) usually need to form metal-ceramic composites with ionic conductors (*e.g.*, GDC) to facilitate ion transport [124]. In contrast, high-entropy oxides (HEOs) be directly used as single-phase anodes due to their mixed ionic-electronic conductivity. Traditional Ni-based anodes suffer from sulfur poisoning and carbon deposition, while perovskite anodes often face insufficient electrical conductivity. The medium-entropy perovskite material $SrV_{1/3}Fe_{1/3}Mo_{1/3}O_3$ forms a disordered structure and exhibits excellent chemical stability in a reducing atmosphere. The small-polaron pair mechanism enhances its conductivity, and the porous structure meets the requirements for anode current collection. Additionally, the high oxygen vacancy concentration improves catalytic activity, enabling a peak power density of 720 mW·cm$^{-2}$ at 850 °C when using $La_{0.9}Sr_{0.1}Ga_{0.8}Mg_{0.2}O_{3-x}$(LSGM) electrolyte [125]. The high-entropy spinel $(CuLiFeCoNi)_{1.4}Mn_{1.6}O_{4-\delta}$ also exhibits good structural and thermal stability. The enhanced Lewis acid-base sites contribute to its superior catalytic activity for both the hydrogen oxidation and oxygen reduction reactions. The area-specific resistance of the symmetric cell is 0.152 Ω·cm$^{-2}$ in air at 600 °C [126]. However, the corresponding fuel cell performance still requires further investigation.

## 5. Conclusions and perspectives

The core strategy of entropy engineering involves incorporating multiple elements into a single-phase solid solution. In SOFCs, high-entropy design can significantly enhance structural stability, suppress cation segregation (such as Ba/Sr), and improve tolerance to contaminants like $CO_2$, $H_2O$, and Cr poisoning. Among the currently reported high-performance high-entropy cathodes, cobalt-based systems still dominate, and oxygen reduction activity and stability can be optimized through A/B-site entropy doping. Notably, Ruddlesden–Popper-type layered structures exhibit intrinsic high-efficiency oxygen transport capabilities, enabling excellent electrochemical performance even in cobalt-free compositions, thus highlighting a promising direction for developing new cathode materials. High-entropy electrolytes show unique advantages due to their good chemical stability and optimized sinterability. And for anode, although still in its early stages, the inherent mixed ionic-electronic conductivity and potential resistance to coking and sulfur poisoning make high-entropy oxides promising candidates for direct hydrocarbon fuel utilization. Several fundamental challenges remain to be addressed before high-entropy oxides can achieve their full potential in SOFC applications.

(a) Composition design and phase formation prediction. Although high-entropy materials offer a wide range of compositional design flexibility, high configurational entropy does not always guarantee the formation of a single-phase solid solution. The applicability of existing phase formation criteria based on conventional materials to high-entropy systems remains unclear. For example, the stable tolerance factor range of high-entropy perovskites differs from that of conventional perovskites, and the permissible threshold for ionic radius differences is higher than that in high-entropy alloy systems. More accurate and reliable predictive design criteria are therefore needed.

(b) High-entropy effect or element inherent properties. Performance improvements in high-entropy materials are often broadly attributed to high-entropy or cocktail effects, but the specific mechanisms require further clarification. For example, enhanced ORR activity may be related to the reduced oxygen vacancy formation energy, but it remains necessary to distinguish whether this improvement stems from increased lattice distortion and dislocations or from the intrinsic contributions of specific elements. Systematic experimental designs, such as replacing specific elements while keeping configurational entropy constant or adjusting entropy while fixing elemental compositions, are needed to identify the respective roles of configurational entropy and intrinsic elemental properties. (c) Rational elemental selection and configurational entropy design. Equimolar designs maximize configurational entropy and favor single-phase formation, but blindly pursuing the highest entropy does not necessarily yield the best performance. A more effective strategy is entropy-tailored modification, such as intentionally introducing specific elements into state-of-the-art materials, using the expanded compositional choice provided by high-entropy engineering to address their performance bottlenecks.

## Acknowledgements

This work was supported by the National Natural Science Foundation of China (Grant Number: 52302263).